# Dynamical Response of Nanomechanical Resonators to Biomolecular Interactions

Kilho Eom[1,*], Tae Yun Kwon[1,2], Dae Sung Yoon[1,†], Hong Lim Lee[2], and Tae Song Kim[1,‡]

[1]*Nano-Bio Research Center, Korea Institute of Science and Technology (KIST), Seoul 136-791, Republic of Korea*

[2]*School of Advanced Materials Science and Engineering, Yonsei University, Seoul 120-749, Republic of Korea*

We studied the dynamical response of a nanomechanical resonator to biomolecular (e.g. DNA) adsorptions on a resonator's surface by using theoretical model, which considers the Hamiltonian $H$ such that the potential energy consists of elastic bending energy of a resonator and the potential energy for biomolecular interactions. It was shown that the resonant frequency shift for a resonator due to biomolecular adsorption depends on not only the mass of adsorbed biomolecules but also the biomolecular interactions. Specifically, for dsDNA adsorption on a resonator's surface, the resonant frequency shift is also dependent on the ionic strength of a solvent, implying the role of biomolecular interactions on the dynamic behavior of a resonator. This indicates that nanomechanical resonators may enable one to quantify the biomolecular mass, implying the enumeration of biomolecules, as well as gain insight into intermolecular interactions

---

[*] E-mail: eomkh@kist.re.kr
[†] Also at Department of Biomedical Engineering, Yonsei University, Kangwondo 220-740, Korea (ROK)
[‡] E-mail: tskim@kist.re.kr

between adsorbed biomolecules on the surface.

Nanomechanical resonators have recently allowed one to not only gain insight into fundamentals of quantum mechanics[1-4] but also detect the molecules even in extremely low concentrations.[5,6] For instance, Yang, et al.[7] reported the ultrahigh sensitive mass sensing of molecules even in a zepto-gram resolution by using a nanomechanical resonator. Moreover, it was recently reported that resonating cantilevers have enabled the sensitive *in-vitro* biomolecular detection.[8-12] The high sensitivity in detecting molecules is attributed to scaling down that leads to high-frequency dynamical range of a resonator. Accordingly, nanomechanical resonators have been a strong candidate for ultrahigh sensitive *in-vitro* biomolecular detection.

The detection principle is the direct transduction of biomolecular adsorption on a resonator's surface into the resonant frequency shift. It was well known that the mass of adsorbed molecules makes contribution to resonant frequency shift,[13] as long as molecular interactions between adsorbed molecules do not play a critical role on the elastic bending behavior of a resonator. In recent studies,[14,15] it was found that resonant frequency shift for *in-vitro* biomolecular detection is ascribed to molecular interactions (e.g. electrostatic repulsion, hydration) between adsorbed biomolecules. Specifically, it was reported that the surface stress induced by biomolecular interactions dominates the resonant frequency shift for *in-vitro* biomolecular detection.[16,17] However, a continnum model with a constant surface stress in recent studies[16,17] may be debatable, since it was provided that, in classical elasticity, the constant surface stress may not induce any resonant frequency shift.[18,19] Moreover, it is hard to quantitatively relate the surface stress to biomolecular interactions. Thus, it is demanded to develop the model based on molecular model of biomolecular interactions for gaining insight into quantitative descriptions on relationship between biomolecular interactions and resonant frequency shift.

In this paper, we developed a model, which allows one to quantitatively describe the role of the intermolecular interactions on the resonance behavior of a nanomechanical resonator, on the basis of the molecular model for biomolecular interactions. Specifically, a model considers the Hamiltonian $H$ for the adsorption of double-stranded DNA (dsDNA) on the surface of nanomechanical resonator such that potential energy includes the elastic bending energy of a resonator and the potential energy for intermolecular interactions between dsDNAs. It was shown that ionic strength of a solvent, which is responsible for intermolecular interactions for dsDNAs, plays a role on the resonant frequency shift. The results allow one to gain insight into not only the relationship between molecular interactions and resonant frequency shift but also how to design the nanomechanical resonator for highly sensitive *in-vitro* biomolecular detection (e.g. DNA detection).

Here, we consider the dynamic behavior of a nanomechanical resonator, which is operated in a NaCl solvent, in response to biomolecular adsorption on its surface. Let us denote the packing density, $\theta$, of adsorbed biomolecules on a surface as $\theta = N/L$, where $N$ is the number of adsorbed biomolecules on a surface and $L$ is a resonator's length. As shown in Fig. 1, once biomolecules are adsorbed on the surface, the intermolecular interaction (e.g. DNA-DNA interaction) induces the additional bending of a resonator. In Fig. 1, the interspacing distance, $d$, between biomolecules (e.g. DNA)

is given by

$$d(s) = d_0\left[1 + \kappa c(1 + s/c)\right] \quad (1)$$

where $d_0 = 1/\theta$, $\kappa$ is a curvature defined as $\kappa = \partial^2 w(x, t)/\partial x^2$ with given deflection $w(x, t)$, $2c$ is a thickness of a resonator, and $s$ is a distance from a resonator's surface.

With the prescribed potential energy, $U(d)$, for intermolecular interactions between adsorbed biomolecules, the effective potential energy, $V$, for a nanomechanical resonator upon biomolecular adsorption on its surface consists of elastic bending energy, $E_b$, of a resonator and potential energy, $E_{int}$, for intermolecular interactions between adsorbed biomolecules.

$$V = E_b + E_{int} = \frac{1}{2}\int_0^L \xi \kappa^2 dx + \int_0^L \theta U dx \quad (2)$$

where $\xi$ is a bending modulus for a resonator. By using Taylor series expansion of $U$ with respect to curvature $\kappa$ at $\kappa = 0$, the total potential energy $V$ is in the form of

$$V = \int_0^L \left[v_0 + \varphi\kappa + (1/2)(\xi + \psi)\kappa^2 + O(\kappa^3)\right]dx \quad (3)$$

Here, the coefficients $v_0$, $\varphi$, and $\psi$ are defined as follows: $v_0 = \theta U|_{\kappa=0}$, $\varphi = \partial(\theta U)/\partial\kappa|_{\kappa=0}$, and $\psi = \partial^2(\theta U)/\partial\kappa^2|_{\kappa=0}$. The kinetic energy, $T$, of a nanomechanical resonator is given by

$$T = \frac{1}{2}\int_0^L (\mu + \theta m)(\partial w/\partial t)^2 dx \quad (4)$$

where $\mu$ is a resonator's mass per unit length, and $m$ is the mass of a biomolecule (e.g. DNA chain). The oscillating deflection motion of a resonator can be represented in the form of $w(x, t) = u(x)\exp[i\omega t]$, where $u(x)$ is a deflection eigenmode and $\omega$ is a resonant frequency. The mean value of Hamiltonian, $<H>$, per oscillation cycle is

$$\langle H \rangle = \langle T \rangle + \langle V \rangle$$
$$= -\frac{\omega^2}{2} \int_0^L (\mu + \theta m) u^2 dx + \int_0^L \left[ v_0 + \varphi u'' + (1/2)(\xi + \psi)(u'')^2 \right] dx \quad (5)$$

where angle bracket < > indicates the mean value per oscillation cycle, and prime represents the differentiation with respect to coordinate $x$. The variational method with a Hamiltonian <$H$> provides the weak form of equation of motion.[20]

$$\delta \langle H \rangle = \int_0^L \left[ -\omega^2 (\mu + \theta m) u + (\xi + \psi)(d^4 u / dx^4) \right] \delta u$$
$$+ \left[ \varphi + (\xi + \psi) u'' \right] \delta u' \Big|_0^L - (\xi + \psi) u''' \delta u \Big|_0^L = 0 \quad (6)$$

Here, a symbol $\delta$ indicates the variation, and one may regard $\delta u$ as a virtual deflection eigenmode that satisfies the essential boundary condition. In Eq. 6, the integrand represents the equation of motion for a resonator with biomolecular adsorptions on its surface, whereas the other terms provide the boundary conditions. Thus, the equation of motion for an oscillating resonator upon biomolecular adsorption on its surface is given by $(\xi + \psi)(d^4u/dx^4) - \omega^2(\mu + \theta m)u = 0$. Consequently, the resonant frequency, $\omega$, of a nanomechanical resonator upon biomolecular adsorptions on its surface is

$$\frac{\omega}{\omega_0} = \sqrt{\frac{1 + (\psi/\xi)}{1 + (\theta m/\mu)}} \quad (7)$$

where $\omega_0$ is a reference resonance, which is a resonance without any biomolecular adsorption, given by $\omega_0 = (\lambda/L)^2 (\xi/\mu)^{1/2}$. As shown in Eq. 7, the resonant frequency shift due to biomolecular adsorption is attributed to not only the mass of adsorbed biomolecules but also the bending stiffness change induced by the intermolecular interactions between adsorbed biomolecules. Specifically, the bending stiffness change induced by biomolecular interactions is dictated by the harmonic (second-order) term $\psi$ in the potential energy for intermolecular interactions. Hence, the resonant frequency shift $\Delta\omega$ due to biomolecular adsorption is represented in the form of

$$\frac{\Delta\omega}{\omega_0} = \frac{\omega - \omega_0}{\omega_0} \approx -\frac{1}{2}\frac{\theta m}{\mu} + \frac{1}{2}\frac{\psi}{\xi} \qquad (8)$$

Here, the negative sign indicates the decrease of resonant frequency after biomolecular adsorption whereas the positive sign represents the increase of resonant frequency after biomolecular adsorption. The first term represents the effect of adsorbed biomolecular mass on the resonant frequency shift while the second term indicates the effect of bending stiffness change induced by intermolecular interactions.

In this work, we consider the case where dsDNA molecules are adsorbed on the surface of a resonator. The intermolecular interactions, $U(d)$, between dsDNAs on the surface was provided by Strey, et al.[21,22]

$$\frac{U(d)}{L_c} = \alpha \frac{\exp(-d/\lambda_H)}{\sqrt{d/\lambda_H}} + \beta \frac{\exp(-d/\lambda_D)}{\sqrt{d/\lambda_D}} + E_{conf}(d) \qquad (9)$$

Here, intermolecular interaction $U$ consists of hydration repulsion with amplitude $\alpha$ and screening length scale $\lambda_H$, electrostatic repulsion with amplitude $\beta$ and Debye length $\lambda_D$, and configurational entropic effect $E_{conf}(d)$ that enhances the hydration and electrostatic repulsions. It should be noted that hydration and electrostatic repulsions are governed by the ionic strength, $[I]$, of a solvent in such a way that the screening lengths and repulsion amplitudes depend on the ionic strength, e.g. $\lambda_D \approx 3.08/\sqrt{[I]}$ Å and $\lambda_H \approx 2.88$ Å for monovalent salt.[22] The packing density of adsorbed DNA molecules is restricted as $0 < \theta \leq 10^9$, because the minimum interspacing distance $d_{0,min}$ for DNA adsorption on cantilever surface is given by $d_{0,min} \sim 1$ nm.[23] From Eqs. 1, 3 and 9, the induced bending stiffness change, $\psi$, for a resonator due to DNA-DNA interactions is computed as $\psi = L_c c^2 f(\theta, [I])$. This indicates that the induced bending stiffness change, $\psi$, depends on geometry parameters for both dsDNA and resonator, i.e. dsDNA chain length $L_c$ and resonator's thickness $2c$. Moreover, in Fig. 2, it is shown that induced bending stiffness change, $\psi$, due to DNA-DNA interactions is dependent on ionic strength of monovalent salt of a solvent, $[I]$, which governs the hydration and electrostatic repulsions, as well as dsDNA packing density. A high packing density and 1M NaCl concentration of a solvent induces the larger repulsive forces between dsDNAs than a low packing density and 0.1 M NaCl concentration of a solvent, resulting in a larger elastic bending motion of a resonator for a high packing density and 1M NaCl concentrations of a solvent. This is consistent with a recent study,[24] which reported that the nanomechanical bending motion of a cantilever is originated from intermolecular interactions between adsorbed biomolecules.

As stated earlier in Eq. 8, the resonant frequency shift due to immobilization of dsDNA on a resonator's surface is determined by both the mass of adsorbed dsDNA molecules and the intermolecular interactions such as hydration and electrostatic repulsion. For understanding such resonant frequency shift, we take into account the nanomechanical cantilever with dimension of $b \times 2c \times L$ (width × thickness × length), where the width and the length are fixed as $b = 200$ nm and $L = 2$ μm. It should be noted that the mass of a single dsDNA chain is given by $m = 10^6$ Dalton (where 1 Dalton = $1.66 \times 10^{-27}$ kg), and that the mass per unit length $\mu$ for a cantilever is given as $4.66 \times 10^{-8}$ g/m, $9.32 \times 10^{-8}$ g/m, and $4.66 \times 10^{-7}$ g/m for a cantilever thickness $2c$ of 100 nm, 200 nm, and 1μm, respectively. Fig. 3 presents the relationship between the resonator's thickness and the resonant frequency shift of a cantilever due to dsDNA adsorption. It is shown that the dimensionless resonant frequency shift, $\Delta\omega/\omega_0$, becomes larger as the thickness is smaller, indicating the role of thickness on the resonant frequency shift.

Furthermore, as shown in Fig. 3, the decrease in the resonant frequency of a cantilever after dsDNA adsorption suggests that the mass of adsorbed dsDNA molecules plays a crucial role on the resonant frequency shift rather than intermolecular interactions. The intermolecular interactions between adsorbed dsDNA molecules play a secondary role on the resonant frequency shift.

In order to gain insight into the role of intermolecular interactions on the resonant frequency shift, we considered the situation where, in the similar spirit to recent experiment by Wu et al.[24], dsDNA molecules are immobilized on a resonator's surface in a solvent with monovalent salt concentration of 0.1 M NaCl, and then the monovalent salt concentration of a solvent is increased to 1M NaCl. Here, we considered a cantilever with dimension of $b \times 2c \times L$ (width × thickness × length), where $b$ = 200 nm and $L$ = 2 μm. As shown in Fig. 4, the change of the monovalent salt concentration of a solvent from 0.1M to 1M induces the increase in the induced bending stiffness change, $\psi$,[24] which leads to the increase in the resonant frequency of a cantilever. Moreover, the resonant frequency shift due to the increase in the monovalent salt concentration of a solvent does not depend on the cantilever's thickness. This can be easily noted that the resonant frequency shift $\Delta\omega$ induced by increase of monovalent salt concentration of a solvent is given by

$$\Delta\omega = \alpha(f_1 - f_0) \tag{10}$$

Here, $f_1$ and $f_0$ are the values of function $f(\theta, [I])$ measured at monovalent salt concentration of $[I]$ = 1 M NaCl and $[I]$ = 0.1 M NaCl, respectively, and $\alpha$ is a constant given as $\alpha = (1/b)(\lambda/L)^2[3/(2\rho LE)]^{1/2}$, where $\rho$ is the cantilever's density. This indicates that the ionic strength of a solvent has also played a role on the resonant frequency shift, and that the cantilever's thickness does not played any role on the resonant frequency shift induced by change of ionic strength of a solvent. It is consistent with previous studies,[23,24] which reported the significant role of ionic strength of a solvent on the nanomechanical bending motion of a cantilever functionalized by dsDNA molecules.

In summary, our model provides that the resonant frequency shift of a nanomechanical resonator depends on not only the mass of adsorbed dsDNA molecules but also the intermolecular interactions. It is shown that the dimensionless resonant frequency shift due to molecular adsorption is related to the thickness in such a way that thinner resonator exhibits the larger dimensionless resonant frequency shift. More remarkably, the ionic strength of a solvent plays a role on the resonant frequency shift of a nanomechanical cantilever such that the increase of monovalent salt concentration from 0.1 M to 1 M induces the accretion in the bending rigidity of a cantilever,[24] resulting in the increase in the resonant frequency of a cantilever. This implies that the nanomechanical resonator may enable one to study the DNA-DNA interactions on the surface. It is proposed that, based on our theoretical model, the nanomechanical resonator may allow for not only enumerating the DNA molecules[11] but also gaining insight into DNA-DNA interactions.

This work was supported by Intelligent Microsystem Center sponsored by the Korea Ministry of Science and Technology as a part of the 21$^{st}$ Century's Frontier R&D projects (Grant No. MS-01-133-01) and the National Core Research Center for Nanomedical Technology sponsored by KOSEF (Grant No. R15-2004-024-00000-0).

**Figure Captions**

Fig. 1. (Color Online) Schematic for a bending of a resonator (e.g. cantilever) induced by intermolecular interactions between adsorbed biomolecules (e.g. DNA)

Fig. 2. (Color Online) Induced bending stiffness change, $\psi/L_c c^2 \equiv f(\theta, [I])$, for a resonator due to dsDNA adsorption was computed as a function of packing density, $\theta$, and ionic strength, $[I]$, of a solvent. Here, normalized dsDNA packing density is given by $\eta = \theta/10^9$.

Fig. 3. (Color Online) Normalized resonant frequency shifts induced by dsDNA adsorption on a surface as a function of resonator's thickness $2c$. Here, we used $[I] = 1$ M NaCl for ionic strength of a solvent, $L_c = 100$ nm for DNA chain length, and the characteristic parameters for a cantilever as follows: $b = 200$ nm, $L = 2$ μm, and $E = 190$ GPa for silicon cantilever.

Fig. 4. (Color Online) Resonant frequency shift for a cantilever functionalized by dsDNA molecules due to the increase in the ionic strength of monovalent salt concentration from 0.1 M to 1 M.

**Figures**

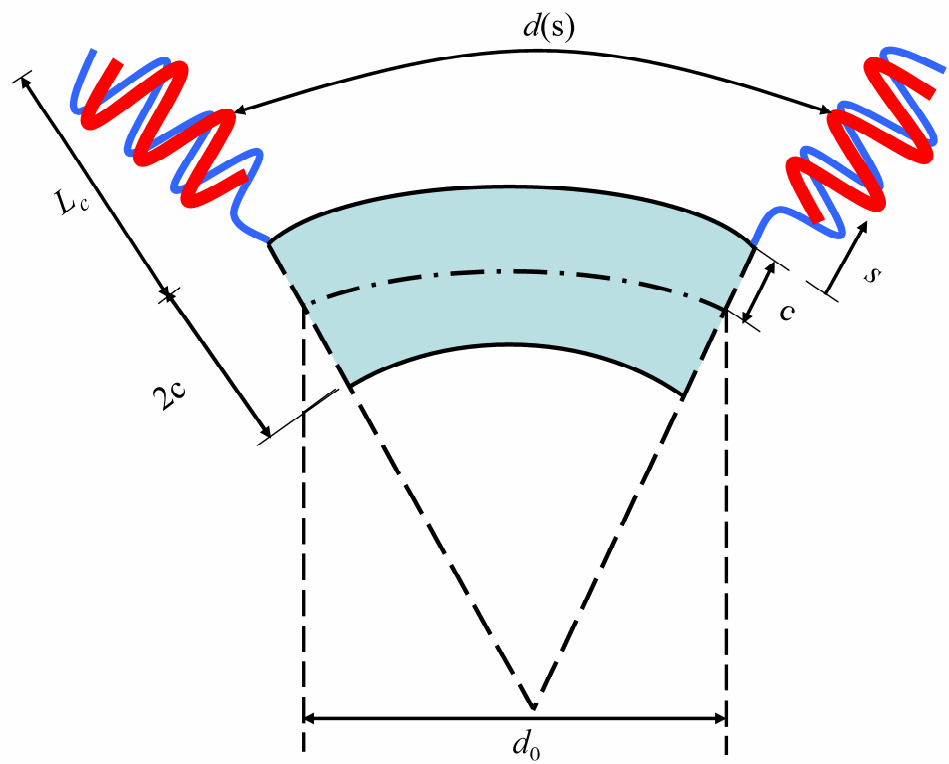

*Fig. 1.* *Eom, K., et al.*

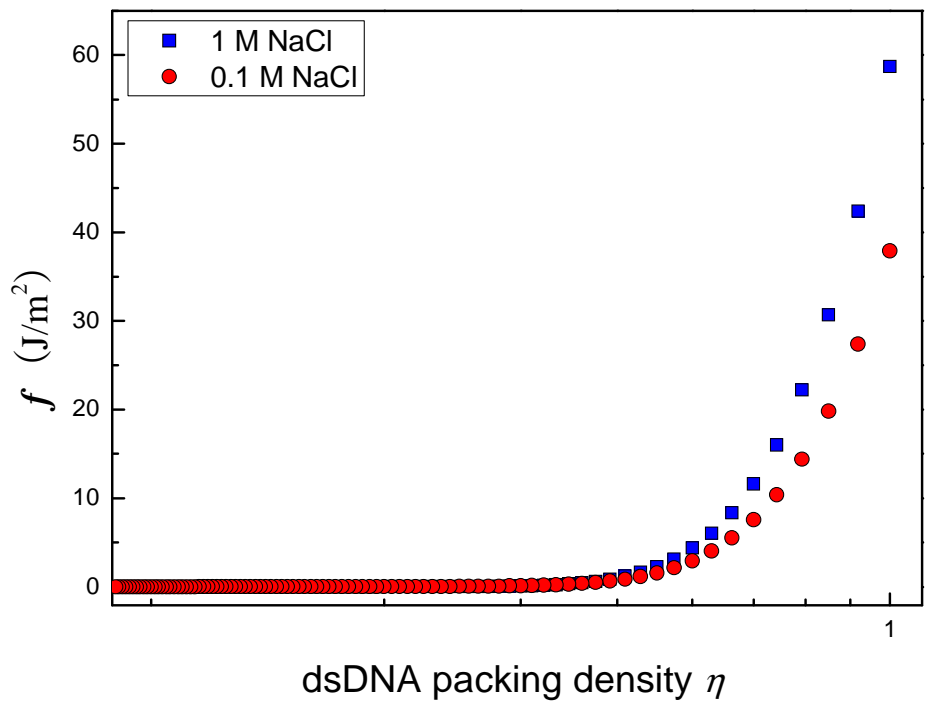

*Fig. 2.* *Eom, K., et al.*

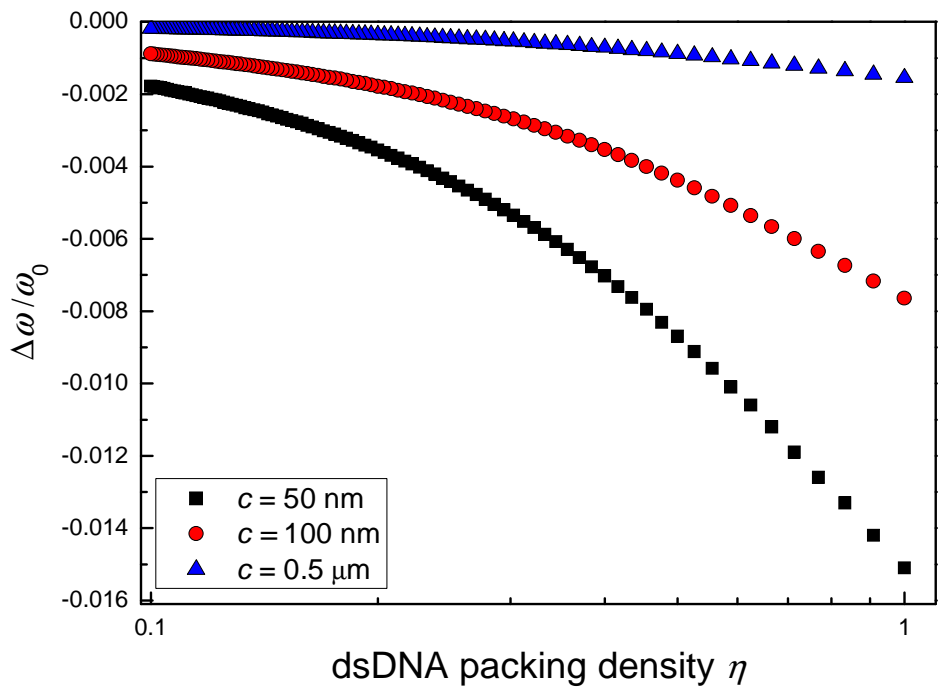

*Fig. 3.*                          *Eom, K., et al.*

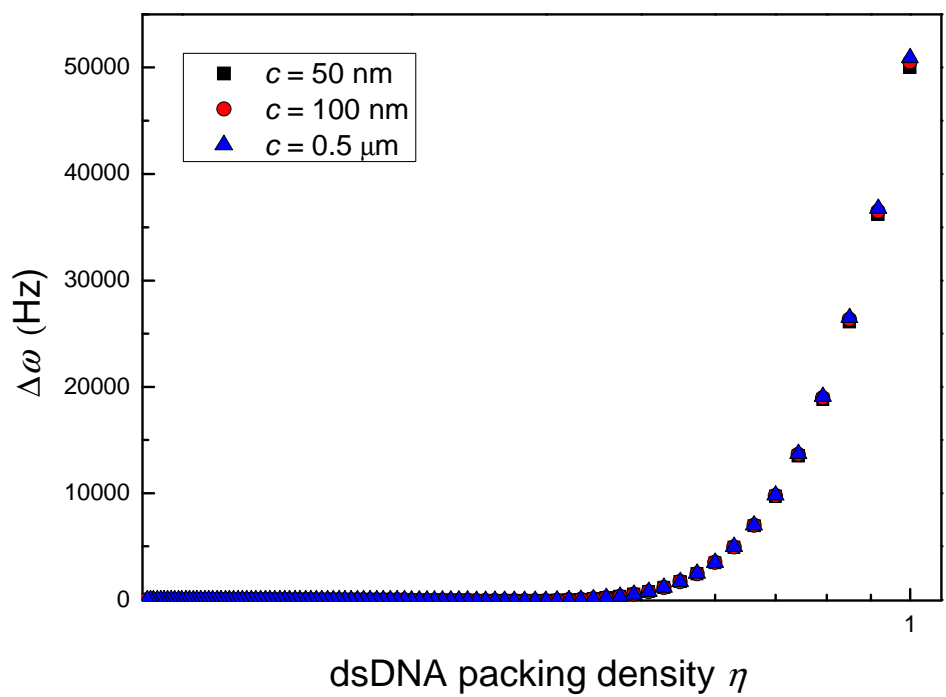

*Fig. 4.*                                                          *Eom, K., et al.*